\documentclass[conference]{IEEEtran}
\IEEEoverridecommandlockouts
\usepackage{cite}
\usepackage{amsmath,amssymb,amsfonts}
\usepackage{algorithmic}
\usepackage{graphicx}
\usepackage{textcomp}
\usepackage{xcolor}
\usepackage{subcaption}
\usepackage{xurl}

\usepackage{hyperref}
\usepackage{balance}
\usepackage{float}
\def\BibTeX{{\rm B\kern-.05em{\sc i\kern-.025em b}\kern-.08em
    T\kern-.1667em\lower.7ex\hbox{E}\kern-.125emX}}

\usepackage{tikz}
\definecolor{brick}{RGB}{178,34,34}
\newcommand{\barval}[1]{%
  \begin{tikzpicture}[baseline=0.5ex]
    \node[anchor=east] at (0,0.125) {#1};
    \begin{scope}[shift={(0.1,0)}] 
      \fill[brick] (0,0) rectangle (#1*0.6, 0.25); 
      \draw[gray!50] (0,0) rectangle (0.6,0.25); 
    \end{scope}
  \end{tikzpicture}%
}

\begin{document}

\title{Can an LLM Detect Instances of Microservice Infrastructure Patterns?\\

\thanks{This work is co-financed by Component 5 - Capitalization and Business Innovation, integrated in the Resilience Dimension of the Recovery and Resilience Plan within the scope of the Recovery and Resilience Mechanism (MRR) of the European Union (EU), framed in the Next Generation EU, for the period 2021 - 2026, within project HfPT, with reference 41.}
}

\author{
    \IEEEauthorblockN{
        Carlos Eduardo Duarte\IEEEauthorrefmark{1},
        Neil B. Harrison\IEEEauthorrefmark{2},
        Filipe Figueiredo Correia\IEEEauthorrefmark{1},
        Ademar Aguiar\IEEEauthorrefmark{1} and
        Pavlína Gonçalves\IEEEauthorrefmark{3}
    }
    \IEEEauthorblockA{
        \IEEEauthorrefmark{1}\textit{INESC TEC}, \textit{Faculdade de Engenharia, Universidade do Porto}, Porto, Portugal
    }
    \IEEEauthorblockA{
        \IEEEauthorrefmark{2}\textit{Department of Computer Science}, \textit{Utah Valley University}, Utah, USA
    }
    \IEEEauthorblockA{
        \IEEEauthorrefmark{3}\textit{INESC TEC}, Porto, Portugal
    }
    \IEEEauthorblockA{
        \{carlos.e.duarte, filipe.correia, ademar.aguiar, pavlina.goncalves\}@inesctec.pt, neil.harrison@uvu.edu
    }
}

\maketitle

\begin{abstract}
Architectural patterns are frequently found in various software artifacts. The wide variety of patterns and their implementations makes detection challenging with current tools, especially since they often only support detecting patterns in artifacts written in a single language. Large Language Models (LLMs), trained on a diverse range of software artifacts and knowledge, might overcome the limitations of existing approaches. However, their true effectiveness and the factors influencing their performance have not yet been thoroughly examined. To better understand this, we developed MicroPAD. This tool utilizes GPT~5~nano to identify architectural patterns in software artifacts written in any language, based on natural-language pattern descriptions. We used MicroPAD to evaluate an LLM's ability to detect instances of architectural patterns, particularly infrastructure-related microservice patterns. To accomplish this, we selected a set of GitHub repositories and contacted their top contributors to create a new, human-annotated dataset of 190 repositories containing microservice architectural patterns. The results show that MicroPAD was capable of detecting pattern instances across multiple languages and artifact types. The detection performance varied across patterns (F1 scores ranging from 0.09 to 0.70), specifically in relation to their prevalence and the distinctiveness of the artifacts through which they manifest. We also found that patterns associated with recognizable, dominant artifacts were detected more reliably. Whether these findings generalize to other LLMs and tools is a promising direction for future research.
\end{abstract}


\section{Introduction}

Software patterns are reusable solutions to recurring problems in a given context~\cite{Gamma1994DesignSoftware}. Patterns can play a significant role in helping practitioners understand how programs work, especially considering that software maintainers spend, on average, around 58\% of their time understanding the software~\cite{Xia2018MeasuringProfessionals}. Notably, preserving the knowledge obtained from \textit{architectural patterns} present in the codebase is essential, as it empowers practitioners to make better decisions related to the system's quality attributes~\cite{Harrison2007LeveragingAttributes}, which in turn influence the system's overall success. Nonetheless, practitioners often fail to maintain software documentation~\cite{Lethbridge2003HowPractice}.

To preserve this knowledge, researchers have proposed several automated tools to identify instances of architectural patterns in software artifacts. However, this task is not simple for several reasons. First, practitioners might not be aware that these patterns exist. Second, they may struggle to identify instances of the patterns across the software artifacts. Third, architectural pattern instances can vary significantly, and their presence can span multiple different artifacts~\cite{Haitzer2015Semi-automaticPrimitives}, making detection more challenging. 

Moreover, existing approaches for detecting architectural patterns are notably limited in several aspects. For example, many existing tools can detect pattern instances only in a single language. Since projects contain artifacts in multiple languages, and architectural patterns can span throughout these artifacts, detection tools may miss many patterns. Another limitation is that the detection mechanisms of existing approaches are specific to the patterns they can detect. Given the wide variety of architectural patterns and the scarcity of available detection tools, there is a need for approaches that enable users to easily detect the specific patterns they require, whether by extending existing tools or creating new ones.

LLMs may address these issues. They have been trained on vast amounts of source code, learning the syntax and semantics of numerous languages while observing many instances of architectural patterns. An LLM-based pattern detector should be able to analyze artifacts in multiple languages without requiring language-specific parsers. However, software repositories are typically too large to process in a single LLM call. Additionally, analyzing every file individually would be prohibitively expensive due to API costs. Consequently, we needed a structured approach to select relevant files and aggregate evidence.

To explore this further, we developed MicroPAD, an automated tool that uses GPT~5~nano to identify architectural patterns from natural-language descriptions. MicroPAD wraps the LLM in a multi-step pipeline that addresses these challenges. Thus, our findings reflect MicroPAD's overall performance rather than the LLM's contribution in isolation. We focused on microservice infrastructure patterns, a subset of microservice architecture patterns defined in Chris Richardson's pattern catalog~\cite{WhatMicroservices}, given the prevalence of microservice architectures in modern software systems and the extensive research interest in this area. Our evaluation across 190 open-source repositories, validated by repository contributors, revealed insights about how pattern characteristics influence detection performance.

The main contributions of this article are the following: 

\begin{itemize} 
    \item MicroPAD, a language-agnostic detection tool requiring only natural language pattern descriptions and repository artifacts as input;
    \item Novel insights on detection performance for microservice architectural patterns, including which pattern characteristics influence detection results;
    \item A human-annotated dataset of 190 repositories with 47 microservice architectural patterns labeled by repository contributors as present or absent in their systems. 
\end{itemize}

These contributions can serve as a basis for further research on LLM-based architectural pattern detection and microservice architectures in general.

\section{Related Work} \label{sec:related-work}

Researchers have proposed various tools for architectural pattern detection using \textit{Heuristics}, \textit{Machine Learning (ML)}, and \textit{LLMs}. Table \ref{tab:relatedwork} summarizes these approaches, detailing their detection techniques, supported languages, dataset sizes, and performance metrics (precision (P), recall (R), accuracy (A), and F1 score (F1)).

\begin{table*}[ht!]
\raggedright
\renewcommand{\arraystretch}{1.5}
\begin{tabular}{p{2.5cm} p{1cm} p{5.4cm} l r r r r r}
\textbf{Work} & \textbf{Technique} & \textbf{Patterns} & \textbf{Language} & \textbf{Sample Size} & \textbf{P} & \textbf{R} & \textbf{A} & \textbf{F1} \\
\hline
Daoudi et al.~\cite{Daoudi2019AnApps} & Heuristics & \textsc{MVW} & Bytecode & 100 & 0.86 & 0.91 & 0.90 & 0.88 \\
Haitzer and Zdun~\cite{Zdun2007SystematicAnalysis} & Heuristics & Any & - & - & - & - & - & - \\
Daniel et al.~\cite{Daniel2023TowardsMetrics} & Heuristics & \textsc{Database Per Service}, \textsc{Single Service Per Host}, \textsc{API Composition}, \textsc{Asynchronous Message}, \textsc{Command-Query Responsibility Segregation} & SQL & - & - & - & - & - \\
Milhem et al.~\cite{Milhem2019ExtractionQualities} & ML & \textsc{Broker}, \textsc{Observer/Publish-Subscribe}, \textsc{Layered Architecture}, \textsc{Pipes and Filters}, \textsc{Shared Repository} & Java &  2 & - & - & - & - \\
Chekhaba et al.~\cite{Chekhaba2021Coach:Apps} & ML & \textsc{MVW} & Java & 265 & - & - & - & 0.56-0.68 \\
Komolov et al.~\cite{Komolov2022TowardsApproach} & ML & \textsc{MVW} & Java & 5,973 & 0.83 & 0.83 & 0.83 & 0.83 \\
Jánki and Bilicki~\cite{Janki2023Rule-BasedModels} & LLMs & \textsc{MVW} & TypeScript & 18,830 & - & - & 0.90 & - \\
Rukmono et al.~\cite{Rukmono2024DeductivePrompting} & LLMs & \textsc{Layered Architecture} & Java & 1 & 0.71 & 0.69 & - & 0.69 \\ 
\hline
\end{tabular}
\caption{A summary of existing tools for detecting architectural patterns.}
\label{tab:relatedwork}
\end{table*}

\subsection{Approaches Based On Heuristics}

Various tools detect instances of architectural patterns by looking, for example, for the presence of particular keywords or calculating code metrics related to each specific pattern~\cite{Daniel2023TowardsMetrics, Daoudi2019AnApps, Haitzer2015Semi-automaticPrimitives}.

\textbf{Daniel et al.}~\cite{Daniel2023TowardsMetrics} present one such tool, which is capable of detecting several microservice architectural patterns. To detect the patterns, the approach analyzes SQL scripts and calculates pattern-related metrics, such as the number of operations exposed by a microservice.

\textbf{Daoudi et al.}~\cite{Daoudi2019AnApps} detect \textsc{Model-View-Whatever (MVW)} patterns (which include patterns such as \textsc{Model-View-Controller (MVC)}, \textsc{Model-View-ViewModel (MVVM)}, and \textsc{Model-View-Presenter (MVP)}) in Dalvik bytecode using structural and event-based heuristics. This approach is difficult to extend, as supporting new patterns requires manually defining new heuristics.

\textbf{Haitzer and Zdun}~\cite{Haitzer2015Semi-automaticPrimitives} overcome single-language limitations by using UML and a Domain-Specific Language (DSL) for pattern detection. However, the approach is semi-automatic and constrained by the DSL's capabilities and learning curve.

\subsection{Approaches Based On ML}

ML approaches~\cite{Zakurdaeva2020DetectingLearning, Milhem2019ExtractionQualities, Komolov2022TowardsApproach, Chekhaba2021Coach:Apps} require training models on pattern-specific datasets, making it costly to extend them to new patterns.

\textbf{Komolov et al.}~\cite{Komolov2022TowardsApproach} use nine ML models to detect \textsc{MVW} patterns in Java programs based on code metrics. However, extending the tool to new languages or patterns requires costly and time-consuming retraining.

\textbf{Chekhaba et al.}~\cite{Chekhaba2021Coach:Apps} use several ML classifiers to detect \textsc{MVW} patterns in Java Android apps. Like Komolov et al., this approach is limited to a single language and requires retraining for expansion.

\textbf{Milhem et al.}~\cite{Milhem2019ExtractionQualities} extend Archie~\cite{Mirakhorli2014Archie:Code} to detect five architectural patterns in Java using ML. Their approach functions by identifying pattern-related terms directly in the source code.

\subsection{Approaches Based On LLMs}

LLMs are significantly impacting how we develop software~\cite{Hou2024LargeReview}. Therefore, it is natural that researchers have begun experimenting with them to identify instances of architectural patterns in software artifacts.

\textbf{Rukmono et al.}~\cite{Rukmono2024DeductivePrompting} employ GPT~4 to detect the \textsc{Layered Architecture} pattern via prompting. The approach is rigid, as the prompts are tailored to a single pattern, offering limited insight into general architectural detection capabilities.

\textbf{Jánki and Bilicki}~\cite{Janki2023Rule-BasedModels} use rule-based prompting to detect \textsc{MVW} patterns in Angular/TypeScript projects. However, defining these rules is labor-intensive, and the approach remains limited to a single language despite using natural language definitions.

\vspace{12pt}

Overall, existing approaches for detecting architectural pattern instances have limitations with respect to the languages they can identify patterns in and the extent to which their implementation depends on the patterns they target. 

Table \ref{tab:relatedwork} shows that most contributions do not disclose any detection performance metrics for the approaches they present, such as F1 scores, making it difficult to compare the detection performance of the different tools. 

Finally, although practitioners are beginning to explore the use of LLMs to detect architectural patterns, the effectiveness of these models in identifying them and their numerous variations remains unclear and warrants further research. In particular, existing LLM-based approaches are limited to specific patterns or languages and require custom prompt engineering.

\section{Research Questions}

Our work employs MicroPAD to automatically detect a set of microservice infrastructure patterns. This evaluation also provides initial insights into how an LLM can detect these patterns.

Our research aims to address two key research questions:

\textbf{RQ1.} \textit{What is the detection performance of a language-agnostic, LLM-based approach to detect microservice infrastructure pattern instances in software artifacts?}

We measured precision, recall, accuracy, and F1 score for each of the nine patterns from Richardson's catalog. Noticing that detection performance varied by pattern, we investigated how it is influenced by the specific file types provided to the LLM. Thus, we formulated the following research question:

\textbf{RQ2.} \textit{How do the artifacts in software repositories impact the ability of an LLM-based approach to detect microservice infrastructure pattern instances?}

We explore these questions using the MicroPAD tool with GPT~5~nano, which will be described in the next section.

\section{Methodology}

To answer these research questions, we first developed a tool to detect instances of architectural patterns in an automated manner using an LLM. We then obtained a set of GitHub repositories and analyzed them with the detection tool. To establish ground truth for evaluating our tool, we conducted a questionnaire survey with the top contributors of those repositories to identify which patterns they contained. Finally, we compared the LLM-based approach to the ground truth. The tool, human-annotated dataset, and other materials from the study are available in an empirical package~\cite{MicroPADPackage}.

\subsection{MicroPAD} \label{sec:tool-name}

Software repositories are typically too large to analyze in a single LLM call, so MicroPAD uses a structured pipeline to select relevant files and aggregate evidence across them. MicroPAD is a custom Python pipeline that orchestrates LLM calls through a multi-step process specifically designed for pattern detection. It supports various LLMs and Small Language Models (SLMs) to identify pattern instances in software artifacts, taking a codebase and natural-language descriptions of desired patterns as input. The tool detects pattern instances across languages by processing various textual artifacts, including natural language and source code, without requiring language-specific parsers. 

Users can configure several parameters, such as selecting the language model to use, with options including OpenAI LLMs or SLMs available through the Ollama platform. Additionally, users can specify the number of files to analyze for each pattern and adjust the temperature settings, if supported by the model. New patterns can be added via natural language descriptions without retraining, and the tool provides reasoning traces showing which files contributed to each decision. We have configured the tool with sensible defaults, as described in the next section.

\begin{figure*}[h!]
    \centering
    \includegraphics[width=1\linewidth]{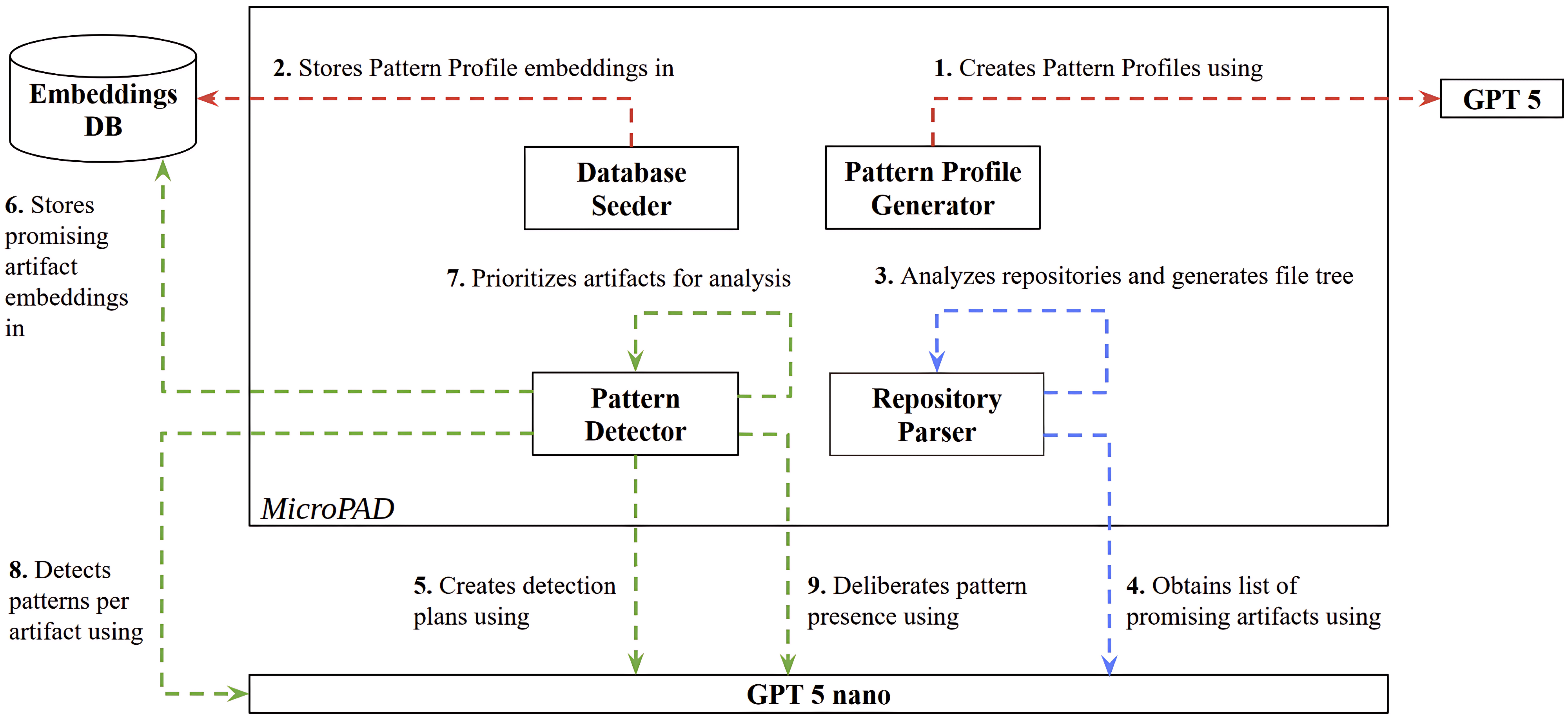}
    \caption{An overview of MicroPAD's main components and interactions.}
    \label{fig:architecture}
\end{figure*}

MicroPAD builds on an earlier prototype~\cite{Duarte2025AutomatedModels} that detected patterns using only Infrastructure-as-Code artifacts. Our approach has four main components: \textbf{Pattern Profile Generator}, \textbf{Database Seeder}, \textbf{Repository Parser}, and \textbf{Pattern Detector}. The \textbf{Pattern Profile Generator} is responsible for creating Pattern Profiles. These are files that contain valuable information to identify pattern instances. The \textbf{Database Seeder} generates and stores embeddings of pattern examples from the Pattern Profiles in a vector database. The \textbf{Repository Parser} filters files likely to contain instances of microservice patterns. Finally, the \textbf{Pattern Detector} identifies pattern instances within the filtered artifacts. 

Figure \ref{fig:architecture} shows MicroPAD's main components and how they interact. The tool executes in three phases: \textit{Creating Pattern Profiles}, \textit{Determining File Paths to Analyze}, and \textit{Detecting Patterns}. Arrows indicate interactions between components, with colors marking each phase: red for \textit{Creating Pattern Profiles}, blue for \textit{Determining File Paths to Analyze}, and green for \textit{Detecting Patterns}.

\subsubsection{Creating Pattern Profiles}

MicroPAD detects patterns using Pattern Profiles. Specifically, these are YAML files containing the pattern's name, description, glob patterns for file path matching (e.g., \textit{**/src/*.py}), and example instances. The LLM uses these glob patterns along with the pattern description to identify relevant files at runtime. To introduce a new pattern, users provide a pattern name, description, and a link to its catalog entry, and MicroPAD uses GPT~5 to generate the Pattern Profile. We manually inspected the generated profiles and verified that glob patterns and examples were reasonable for each pattern.

We chose the following patterns from Richardson's catalog: \textsc{Single Service Instance Per Host}, \textsc{Multiple Service Instances Per Host}, \textsc{Service Instance Per Container}, \textsc{Service Mesh}, \textsc{Service Instance Per VM}, \textsc{Service Deployment Platform}, \textsc{Server-Side Service Discovery}, \textsc{3rd Party Registration}, and \textsc{Service Registry}.

Finally, the \textbf{Database Seeder} creates embeddings from the example instances in the Pattern Profiles and stores them in the ChromaDB vector database.

\subsubsection{Determining File Paths to Analyze}

The \textbf{Repository Parser} scans the repository and generates its file tree, which may be trimmed if too large. It is then sent to the LLM along with the Pattern Profiles. The LLM returns a filtered list of file paths it believes are likely to contain instances of the pattern, together with a confidence score.

This information is used in the next phase to detect pattern instances in the filtered artifacts.

\subsubsection{Detecting Patterns}

The \textbf{Pattern Detector} performs pattern detection in four steps. 

First, during the \underline{Planning Step}, the tool devises a plan to detect each pattern. Second, during the \underline{File Prioritization Step}, the tool determines which artifacts to analyze to minimize costs. Files are selected based on the similarity between Pattern Profile example embeddings and artifact embeddings, the presence of keywords in the artifacts, and the LLM's confidence scores that were calculated during the file tree analysis.

Third, the \textbf{Pattern Detector} moves to the \underline{Investigation Step}. During this step, the tool selects a subset of files with the highest confidence scores for analysis by the LLM. It also sends keywords relevant for detecting the pattern, a repository summary, the detection plan, a pattern description, and an example instance. The LLM analyzes these files to determine which contain evidence of the pattern, offering reasoning, supporting code snippets, and additional relevant information. It also gives a confidence level based on the actual content of the files.

Fourth, the \underline{Deliberation Step} begins. The \textbf{Pattern Detector} reviews summaries of the evidence collected earlier. This includes file paths, confidence scores, and explanations for why patterns may be present. This step is vital as architectural patterns often span multiple files, making individual evidence insufficient. After analysis, the \textbf{Pattern Detector} assigns a final confidence score from 0 to 10 for each identified pattern, requiring a score that exceeds a predefined threshold to confirm its presence.

\subsection{Experiment Configuration} \label{sec:evaluation-method}

We configured the tool with default parameters to recognize real-world patterns, balancing time and cost. For the experiment, we used GPT~5~nano, accessed through OpenAI's API, known for its reasoning capabilities in coding and multi-step planning~\cite{ReasoningAPI}. The tool processed the top 20 files with the highest confidence scores and truncated any files over 50,000 characters to comply with OpenAI's rate limits, while also keeping costs and execution time under control.

For artifact prioritization, we assigned a 70\% weight to the LLM's file-tree analysis to understand how the LLM detects patterns. The remaining 30\% was initially intended to be split equally between keyword matching and embedding similarity. However, due to the limited examples in the Pattern Profiles, we allocated 20\% to keyword matching and 10\% to embedding similarity, using multiples of 10 for simplicity.

We set the confidence score threshold at 5/10 and limited LLM explanations during the Deliberation Step to 220 characters. These and other configuration details are available in the empirical package~\cite{MicroPADPackage}.

\subsection{Repository Selection} \label{sec:obtaining-repos}
To evaluate the tool, we assembled a set of repositories containing examples of microservice infrastructure patterns. Using the APIs of both GitHub and GitHub Archive~\cite{GHArchive}, we identified all repositories with activity on September 30, 2025, between 00:00 and 01:00 GMT, and on October 15, 2025, between 00:00 and 01:00 GMT. We required each repository to have at least ten stars, be active for at least six months, weigh between 100 KB and 100 MB, contain at least three artifacts that matched the glob patterns present in Pattern Profiles, have at least five commits in the past three months, and have a minimum of two contributors.

\subsection{Questionnaire Survey}

To establish ground truth and verify whether the chosen repositories contain the patterns we aimed to identify, we contacted the top two contributors in each repository, based on their activity over the last 100 commits, and invited them to participate in a survey. The survey was delivered through our institution's LimeSurvey platform. To comply with applicable data protection regulations, we ensured that no contributor was contacted more than once.

Although our study focuses on nine microservice infrastructure patterns, we asked practitioners about the presence of 47 patterns from Richardson's catalog. In each pattern identification question section, practitioners had access to a brief description of each pattern.

We collected several demographic data points from the participants, such as their country of residence. Additionally, we asked them how many years of experience they had in professional software development, the approximate number of years they had used microservices in software projects, how familiar they were with microservice pattern concepts, their self-assessed competence in developing microservice-based systems, and their understanding of the software system.

\subsection{Data Analysis}

To answer \textbf{RQ1}, the contributions from respondents, along with the results of MicroPAD, were used to calculate confusion metrics and derive additional statistics, specifically precision, recall, accuracy, and the F1 score.

To answer \textbf{RQ2}, we examined how MicroPAD interacts with the artifacts present in the repositories. To better understand which files are most frequently involved in MicroPAD's detection decisions, we introduced the File Dominance Index (FDI). The FDI is a post-hoc metric computed from MicroPAD's processing logs across all 190 repositories and measures how often MicroPAD analyzes a file compared to the average file. We define it as follows:

\[
  \text{FDI}_i = \frac{c_i}{\mu} = \frac{c_i \cdot N}{T}
\]

\noindent where $c_i$ is the occurrence count of file $i$, $N$ is the total number of unique filenames analyzed for the pattern, $T = \sum_{j=1}^{N} c_j$ is the total file occurrences, and $\mu = T/N$ is the average occurrence count per unique filename.

\section{Results}

We cloned microservice infrastructure repositories, conducted a questionnaire survey, and analyzed the data to answer our research questions.

\subsection{RQ1. Pattern Detection Performance}

The first research question was addressed by comparing the pattern-detection results from our tool with those from repository contributors in the questionnaire survey, which are part of the dataset we built. To the best of our knowledge, no published human-annotated dataset of repositories containing instances of microservice architectural patterns, as defined in Richardson's catalog, exists to date. 

\subsubsection{Dataset Participant Characterization}

We received 208 responses from 4,082 contacted contributors across 48 countries. Most participants had over 8 years of experience. Furthermore, 78.4\% understood microservice patterns, 54.8\% could build them, and 85.6\% were highly familiar with the repositories they contributed to.

\subsubsection{Dataset Repository Characterization}

The participants' 190 annotated repositories form our final dataset. Spanning 35 primary languages (mostly TypeScript, Python, Rust, Go, and C\#), the repositories vary widely in size, popularity, and age, as detailed in Table~\ref{tab:repo_stats}.

\begin{table}[ht]
  \centering
  \renewcommand{\arraystretch}{1.5}
  \begin{tabular}{lrrrrr}
  \textbf{Metric} & \textbf{Min} & \textbf{Max} & \textbf{Mean} & \textbf{Median} & \textbf{SD} \\
  \hline
  Age (years) & 0.7 & 15.2 & 4.8 & 3.8 & 3.7 \\
  Contributors & 1.0 & 8.0 & 1.7 & 1.0 & 1.2 \\
  Size (MB) & 0.1 & 95.7 & 22.9 & 13.8 & 25.4 \\
  Stars & 10.0 & 78,162.0 & 3,132.0 & 352.0 & 8,223.0 \\
  \hline
  \end{tabular}
  \caption{Characterization of the 190 analyzed repositories.}
  \label{tab:repo_stats}
\end{table}

As detailed in Table \ref{tab:pattern_performance} (PV column), pattern prevalence, according to the contributors, varied significantly, ranging from 10\% to 35\%.

\begin{table*}[ht]
  \centering
  \renewcommand{\arraystretch}{1.5}
  \begin{tabular}{lcccccr}
  \textbf{Pattern} & \textbf{PV} & \textbf{P} & \textbf{R} & \textbf{A} & \textbf{F1} & \textbf{Max FDI} \\
  \hline
  \textsc{Service Instance Per Container}      & \barval{0.35} & \barval{0.68} & \barval{0.71} & \barval{0.78} & \barval{0.70} & 54.59 \\
  \textsc{Single Service Instance Per Host}    & \barval{0.32} & \barval{0.54} & \barval{0.11} & \barval{0.68} & \barval{0.19} & 19.61 \\
  \textsc{Multiple Service Instances Per Host} & \barval{0.30} & \barval{0.44} & \barval{0.50} & \barval{0.66} & \barval{0.47} & 48.32 \\
  \textsc{Service Deployment Platform}         & \barval{0.24} & \barval{0.40} & \barval{0.43} & \barval{0.71} & \barval{0.42} & 79.12 \\
  \textsc{Service Instance Per VM}             & \barval{0.17} & \barval{0.75} & \barval{0.09} & \barval{0.84} & \barval{0.16} & 9.15 \\
  \textsc{Service Registry}                    & \barval{0.14} & \barval{0.29} & \barval{0.07} & \barval{0.84} & \barval{0.12} & 3.67 \\
  \textsc{Service Mesh}                        & \barval{0.11} & \barval{0.40} & \barval{0.10} & \barval{0.88} & \barval{0.15} & 5.42 \\
  \textsc{3rd Party Registration}              & \barval{0.11} & \barval{0.33} & \barval{0.05} & \barval{0.89} & \barval{0.09} & 1.93 \\
  \textsc{Server-Side Service Discovery}       & \barval{0.10} & \barval{0.27} & \barval{0.22} & \barval{0.87} & \barval{0.24} & 11.56 \\
  \hline
  \end{tabular}
  \caption{Various pattern-related metrics, including: prevalence of each pattern in the repositories, according to the repository contributors (PV); the tool's precision (P), recall (R), accuracy (A), and F1 score (F1), per pattern; and the maximum file dominance index (Max FDI) value, per pattern.}
  \label{tab:pattern_performance}
\end{table*}

\subsubsection{Pattern Detection}

The confusion matrix in Table~\ref{tab:confusion_matrix} enables us to evaluate MicroPAD's ability to recognize microservice infrastructure architectural patterns. It details the number of true positives (TP), false positives (FP), true negatives (TN), and false negatives (FN).

\begin{table}[ht]
\centering
\renewcommand{\arraystretch}{1.5}
\begin{tabular}{lrr}
 & \textbf{Predicted Positive} & \textbf{Predicted Negative} \\
\hline
\textbf{Actual Positive} & 114 (TP) & 234 (FN) \\
\textbf{Actual Negative} & 116 (FP) & 1,246 (TN) \\
\hline
\end{tabular}
\caption{Confusion matrix describing MicroPAD's performance.}
\label{tab:confusion_matrix}
\end{table}

MicroPAD's detection performance varies significantly by pattern (see Table~\ref{tab:pattern_performance}), with F1 scores ranging from 0.70 for \textsc{Service Instance Per Container} to 0.09 for \textsc{3rd Party Registration}. There is a strong positive correlation ($r = 0.74$) between the F1 score and pattern prevalence, showing that MicroPAD effectively detects common patterns such as \textsc{Service Instance Per Container} and \textsc{Service Deployment Platform}, while less prevalent patterns remain more challenging to detect. Overall, \textbf{MicroPAD achieved 79.5\% accuracy, 49.6\% precision, 32.8\% recall, and an F1 score of 39.5\%}. We investigate the factors behind this variation in Section~\ref{sec:discussion}.

\subsection{RQ2. Impact of Repository Artifacts on Detection}

In \textbf{RQ2}, we further examined how the files analyzed by MicroPAD relate to the overall detection scores for each pattern. Table \ref{tab:fdi} shows the top three most frequently analyzed files per pattern, ranked by maximum FDI.

\begin{table}[ht]
  \centering
  \small
  \begin{tabular}{lrr}
  \textbf{Pattern / File} & \textbf{Count} & \textbf{FDI} \\
  \hline
  \textit{Service deployment platform} & & \\
  \quad Makefile & 120 & 79.12 \\
  \quad main.yml & 44 & 29.01 \\
  \quad tsconfig.json & 40 & 26.37 \\
  \hline
  \textit{Service instance per container} & & \\
  \quad Dockerfile & 89 & 54.59 \\
  \quad HEAD & 32 & 19.63 \\
  \quad README.md & 29 & 17.79 \\
  \hline
  \textit{Multiple service instances per host} & & \\
  \quad Makefile & 66 & 48.32 \\
  \quad release.yaml & 27 & 19.77 \\
  \quad Dockerfile & 24 & 17.57 \\
  \hline
  \textit{Single Service Instance per Host} & & \\
  \quad versions.tf & 33 & 19.61 \\
  \quad \_platform\_variables.tf & 32 & 19.02 \\
  \quad project.tf & 32 & 19.02 \\
  \hline
  \textit{Server-side service discovery} & & \\
  \quad \_\_init\_\_.py & 14 & 11.56 \\
  \quad service.yaml & 13 & 10.73 \\
  \quad README.md & 11 & 9.08 \\
  \hline
  \textit{Service instance per VM} & & \\
  \quad 2024-07-01.xml & 205 & 9.15 \\
  \quad 2023-09-01.xml & 196 & 8.75 \\
  \quad 2024-10-01-preview.xml & 193 & 8.61 \\
  \hline
  \textit{Service mesh} & & \\
  \quad mod.rs & 7 & 5.42 \\
  \quad deployment.yaml & 4 & 3.10 \\
  \quad ingress.yaml & 4 & 3.10 \\
  \hline
  \textit{Service registry} & & \\
  \quad mod.rs & 4 & 3.67 \\
  \quad plugin.ex & 3 & 2.75 \\
  \quad Cargo.toml & 3 & 2.75 \\
  \hline
  \textit{3rd Party Registration} & & \\
  \quad docker-compose.yaml & 2 & 1.93 \\
  \quad external\_registry.cpp & 2 & 1.93 \\
  \quad servicediscovery.md & 1 & 0.96 \\
  \hline
  \end{tabular}
  \caption{Top three dominant files per pattern (FDI).}
  \label{tab:fdi}
\end{table}

\textbf{Specific patterns have dominant files that are analyzed significantly more often than average, where some files are analyzed 50 to 80 times more frequently. We found a strong positive correlation ($r=0.83$) between the F1 scores and the maximum FDI values for each pattern, meaning that patterns with more prominent artifacts are easier to identify.} Conversely, patterns such as \textsc{3rd Party Registration} and \textsc{Service Registry} lack any dominant file, with maximum FDI values below 4, and correspond to the lowest F1 scores.

\section{Discussion}\label{sec:discussion}

Our results show that MicroPAD's detection performance with GPT~5~nano varies significantly across patterns. As a tool designed to make pattern detection accessible through natural language descriptions alone, MicroPAD combines engineering decisions (file prioritization, Pattern Profiles, detection threshold) with LLM capabilities (semantic analysis, pattern recognition). Our findings, therefore, reflect the tool as a whole rather than the LLM in isolation.

\subsection{Pattern Prevalence Correlates With Detection Success} As shown in \textbf{RQ1}, there is a strong correlation between pattern prevalence and detection success. More frequently occurring patterns, such as \textsc{Service Instance Per Container}, are identified more effectively, whereas patterns like \textsc{3rd Party Registration} and \textsc{Service Registry}, which have low prevalence and poor F1 scores, are identified less effectively.

One possible explanation is the representativeness of the training data, as common patterns may be better represented in the repositories on which the LLM was trained. Moreover, prevalence alone does not explain performance: \textsc{Multiple Service Instances Per Host} has similar prevalence to \textsc{Service Instance Per Container} but a much lower F1 score, indicating that other factors also affect detection.

\subsection{Artifact Distinctiveness Predicts Detection Success}

The Pearson Correlation Coefficient between the F1 score and FDI indicates a strong positive relationship, revealing that patterns with higher file dominance are detected more successfully. The top-performing patterns are associated with popular technologies like Docker, Terraform, and Kubernetes, which reinforces their strong association in practice. 

However, high file dominance does not guarantee detection success, as seen with the \textsc{Service Deployment Platform} pattern. Despite Makefiles having the highest FDI, this pattern achieves only a modest F1 score. Makefiles are ubiquitous across repositories, regardless of whether a pattern is present, making them a poor discriminator. This indicates that artifact distinctiveness, not just frequency, matters for detection performance.

Conversely, patterns like \textsc{Service Registry} typically manifest through technology-specific artifacts such as \textit{Consul} or \textit{etcd} configurations (see Table~\ref{tab:fdi}). Yet MicroPAD analyzed generic files like \textit{mod.rs} and \textit{Cargo.toml} instead, suggesting that poor file selection also contributes to low detection performance.

\subsection{Comparison With Existing Approaches}

Direct comparison with existing tools (listed in Table~\ref{tab:relatedwork}) is difficult, as only three report F1 scores alongside sample sizes. Moreover, these tools typically detect patterns in a single language, focus on MVW patterns with predictable structures, and require substantial effort to extend. MicroPAD supports multiple languages, targets patterns with high implementation variability, and can be extended by adding a new Pattern Profile without retraining or developing new heuristics. The difference in pattern complexity and scope means that F1 scores are not directly comparable across these approaches.

\subsection{Practical Implications}

For practitioners, MicroPAD is most reliable for patterns with well-defined artifacts such as containers and deployment platforms. Results for patterns with low FDI should be verified manually. MicroPAD is available in the replication package and can be used to explore unfamiliar codebases that lack architecture documentation or to support the creation of such documentation for existing systems. For researchers, the Pattern Profile mechanism enables extending MicroPAD to other pattern catalogs, and FDI may serve as a predictor of which patterns are amenable to automated detection.

\section{Threats to Validity}

We identified several threats to validity and limitations that warrant explicit disclosure, along with the steps we took to minimize them.

\subsection{Construct Validity}

\textbf{Detection threshold.} Different thresholds can yield varying results. We chose 5/10 as the midpoint of the confidence scale to avoid biasing detection toward either over- or under-detection. Preliminary tests with values above 5 confirmed this choice, as they yielded a greater imbalance between precision and recall. A full sensitivity analysis is left for future work.

\textbf{Patterns are subjective constructs.} Practitioners and LLMs may interpret the same pattern definition differently. This is an inherent challenge of architectural pattern detection rather than a limitation of our study, but it does introduce measurement variability. The large sample sizes across repositories, participants, and patterns help absorb this variability.

\textbf{Dataset Imbalance.} Our ground truth dataset is skewed toward true negatives, artificially inflating baseline accuracy. We mitigate this evaluation threat by relying on precision, recall, and F1 scores instead.

\subsection{Internal Validity}

\textbf{Pattern Profile quality.} Pattern Profiles were generated by GPT~5 and manually inspected to verify that glob patterns and examples were reasonable for each pattern. Profile quality may still vary across patterns, as some patterns admit many implementation variations that a limited set of examples cannot fully cover. However, the impact is partially limited, as pattern examples influence only 10\% of the final decision through embedding-based file prioritization.

\textbf{Information scoping.} MicroPAD limits the information sent to the LLM. This is done by truncating files exceeding 50,000 characters, trimming file trees for large repositories, and analyzing only the top 20 files per pattern. These decisions trade completeness for feasibility, since each LLM call incurs API costs. In practice, most infrastructure-relevant files (Dockerfiles, YAML) fall well below the truncation limit, and the file prioritization step compensates for tree trimming by scoring files based on content-level keyword matching and embedding similarity.

\textbf{File selection accuracy.} We did not separately validate the accuracy of MicroPAD's file selection step. The strong correlation between FDI and detection success ($r=0.83$) suggests that file selection performs well for patterns with distinctive artifacts. For patterns without such artifacts, file selection is inherently more challenging regardless of the approach, and improving it is a direction for future work.

\subsection{External Validity}

\textbf{Limited pattern selection.} Resource constraints limited our scope to nine infrastructure patterns from Richardson's catalog. These patterns span diverse implementation strategies, from container deployment to service discovery and registration, providing meaningful variation in detection difficulty.

\textbf{Generalizability.} The 190 analyzed repositories are open-source projects from GitHub; industrial or private repositories may differ. Additionally, results characterize MicroPAD with GPT~5~nano specifically, and we intentionally scope our claims to this tool rather than to LLMs in general. Whether these findings extend to other LLMs, tools, or agentic approaches remains an open question for future research. We make all data and code available to enable such comparative studies.

\subsection{Conclusion Validity}

\textbf{LLM non-determinism.} Since GPT~5~nano does not allow setting its temperature to zero, we ensured reproducibility by using predefined seeds, logging exact model versions and timestamps, and open-sourcing all code and prompts.

\section {Conclusion and Future Work}

In this article, we presented three main contributions. First, an extensible LLM-based tool that automates the detection of microservice architectural patterns, paving the way for broader software pattern detection tools. Second, we provide key insights into the automated detection of microservice infrastructure patterns, informing future LLM research and identifying characteristics that make patterns detectable. Finally, we introduce a novel dataset of repositories containing microservice architectural patterns to assist practitioners in studying microservice architectures.

Several directions for future work emerge from these findings. Evaluating the approach with additional LLMs would clarify whether the performance patterns we observed are model-specific or reflect broader challenges in architectural pattern detection. Investigating which architectural characteristics make patterns easier or harder to identify could guide both tool development and pattern catalog design. Further improvements to file selection strategies, optimization of the detection threshold through sensitivity analysis, and the extension to additional patterns from Richardson's catalog are also promising directions. Finally, exploring whether richer pattern examples, smaller language models, or fine-tuned LLMs can improve detection performance remains an open question.

\section{Empirical Package}

The annotated data set and MicroPAD's source code are available for replication and further studies on Zenodo \cite{MicroPADPackage} and GitHub \cite{Ceduarte31/micropad:MicroPAD}. To respect the privacy of all involved users, all data containing personal user information has either been anonymized or withheld.

\section*{Acknowledgment}

We sincerely thank all survey participants for their insights. We acknowledge the use of Claude 4.5 Sonnet for code refactoring and documentation, and Grammarly for proofreading the manuscript.

\balance

\bibliographystyle{IEEEtran}
\bibliography{references}

\end{document}